\documentclass[twocolumn,showpacs,preprintnumbers,aps,pra,amsmath,amssymb,floatfix,superscriptaddress]{revtex4-1}

\usepackage{color}
\usepackage{graphicx}      
\usepackage{subfigure}
\usepackage{dcolumn}
\usepackage{textcomp}
\usepackage{longtable}     
\usepackage{url}           
\usepackage{bm}            
\usepackage{appendix}
\usepackage{hyperref}
\usepackage{soul}

\newcommand{\ket}[1]{\left|{#1}\right\rangle}

\begin{document}

\title{Manipulation of heteronuclear spin dynamics with microwave and vector light shift}

\author{Lintao Li}
\affiliation{Department of Physics, The Chinese University of Hong Kong, Hong Kong, China}
\author{Bing Zhu}
\affiliation{Physikalisches Institut, Universit\"at Heidelberg, Im Neuenheimer Feld 226, 69120 Heidelberg, Germany}
\affiliation{Hefei National Laboratory for Physical Sciences at the Microscale and Shanghai Branch, University of Science and Technology of China, Shanghai 201315, China}
\affiliation{CAS Center for Excellence and Synergetic Innovation Center in Quantum Information and Quantum Physics, University of Science and Technology of China, Shanghai 201315, China}
\author{Bo Lu}
\affiliation{School of Physics and Astronomy, Sun Yat-Sen University, Zhuhai 519082, China}
\author{Shizhong Zhang}
\affiliation{Department of Physics and Center of Theoretical and Computational Physics, The University of Hong Kong, Hong Kong, China}
\author{Dajun Wang}
\email{djwang@cuhk.edu.hk}
\affiliation{Department of Physics, The Chinese University of Hong Kong, Hong Kong, China}
\affiliation{The Chinese University of Hong Kong Shenzhen Research Institute, Shenzhen, China}

\begin{abstract}
We report the observation and manipulation of heteronuclear spin dynamics in a spin-1 mixture of ultracold $^{87}$Rb and $^{23}$Na atoms. The dynamics is driven by the interspecies spin-dependent interaction and shows a pronounced dependence on magnetic fields with influences from both linear and quadratic Zeeman shifts. Similar to the well-studied homonuclear cases, the interspecies spin dynamics can be controlled by tuning the quadratic Zeeman shift with far-detuned microwave fields. In addition, we successfully realize spin dynamics control with vector light shifts which act as a species-selective effective magnetic field on $^{87}$Rb atoms. Both methods show negligible loss of atoms thus will be powerful techniques for investigating spin dynamics with fast temporal and high spatial resolutions.
\end{abstract}


\maketitle


\section{Introduction}

The spin degree of freedom in ultracold atomic gases has been recognized as a great asset for rich physics since 1998~\cite{HO1998, OHMI1998}. Indeed, in these so-called ultracold spinor gases, colletive spin dynamics in the form of coherent spin population oscillations have been investigated with Bose condensated atoms~\cite{STENGER1998, KUWAMOTO2004, SCHMALJOHANN2004, CHANG2004, CHANG2005, WIDERA2005, KRONJAGER2006, GERBIER2006, BLACK2007, KLEMPT2009}, as well as thermal Bose gases~\cite{PECHKIS2013, XDHE2015} and degenerate Fermi gases~\cite{KRAUSER2014, EBLING2014}. With the help of these spin dynamics, spin squeezing and multi-particle entanglement with potential applications in quantum metrology and quantum information have been generated~\cite{HAMLEY2012, BOOKJANS201102, HOANG2016, XYLUO2017, ANDERS2018}.

Central to the spin dynamics in spinor gases is the spin-dependent interaction, which favors either ferromagetic (for spin-1 $^{87}$Rb) or antiferromagnetic (for spin-1 $^{23}$Na and spin-2 $^{87}$Rb) phases depending on its sign. In addition, the spin dynamics and ground states in homonuclear spinor systems are also dependent on the quadratic Zeeman energy which compete with the spin-dependent interaction. Although the spin-dependent interaction is tied to the detailed interaction properties of the atom and is thus hard to tune, the quadratic Zeeman energy can be readily manipulated to control the behavior of spinor gases. With homonuclear spinor gases, many interesting physics has been explored along this line by either changing the magnetic field directly or by applying a detuned microwave field~\cite{WIDERA2005, LESLIE2009, BOOKJANS2011, LCZHAO2014}.

Spin dynamics can also occur between different atomic species, e.g. between spin-1 $^{23}$Na and spin-1 $^{87}$Rb. Different from the homonuclear spin-1 case in which there are only two allowed total spin channels with $F = 0$ and 2, total spin channels $F = 0,1,2$ are all allowed in the heteronuclear spin-1 mixture. While spin oscillations in homonuclear spin-1 systems can only happen via the $2\ket{m=0} \leftrightarrow \ket{m=1} + \ket{m=-1}$ process, several processes are allowed in heteronuclear spin-1 systems~\cite{ZFXU2012}. Here $\ket{m=0,\pm 1}$ denote the Zeeman sublevels of the $f = 1$ hyperfine state.   

Previously, we have observed the coherent heteronuclear spin oscillation~\cite{XKLI2015} in a spin-1 mixture of $^{87}$Rb and $^{23}$Na atoms via the spin exchange process $\ket{m_1 = 0,m_2 = -1} \leftrightarrow \ket{m_1 = -1,m_2 = 0}$. Here, following the same notation as~\cite{XKLI2015}, we denote a pair of spin-1 atoms with $^{87}$Rb in $\ket{m_1}$ and $^{23}$Na in $\ket{m_2}$ as $\ket{m_1,m_2}$. While the homonuclear spin dynamics is only sensitive to the quadratic Zeeman shift, heteronuclear spin dynamics depends on the total Zeeman energy difference between the two sides of this process, including contributions from both the linear and quadratic Zeeman shifts. In addition, the heteronuclear spin dynamics is also very sensitive to the ``fictitious magnetic field'' from the vector light shift generated by the optical trap laser in a species-specific manner~\cite{COHEN1972, XKLI2015}.

In this work, we explore in detail the various aforementioned methods for controlling the heteronuclear spin dynamics. We start from a pure $\ket{0,0}$ mixture of $^{87}$Rb and $^{23}$Na atoms and focus on the process (which we call process~(1))
\begin{equation}
\ket{0,0} \leftrightarrow \ket{1,-1}.
\label{eq0}
\end{equation} 
We have observed the spin population oscillation following this process and studied its dependence on magnetic field $B$. We then demonstrate controlling of the spin dynamics by either a detuned microwave field or a laser beam, with both of them selectively dressing the energy levels of $^{87}$Rb atoms.



The paper is organized as follows. In section~\ref{theory} we go over the theories for tuning the quadratic Zeeman shift of spin-1 atoms with a detuned microwave and inducing an effective magnetic field with circular polarized light. In section~\ref{experiment}.A, B and C, we describe the experimental setup and the observation of heteronuclear spin population with both a dual BEC mixture and a thermal $^{87}$Rb $+$ $^{23}$Na BEC mixture. The main results of controlling the spin dynamics with different dressing fields are presented in section~\ref{experiment}.D and \ref{experiment}.E.


\section{Theory}
\label{theory}


\subsection{Heteronuclear spin dynamics in the Rb and Na spin-1 mixture}

In the basis of total spin, the heteronuclear interaction between two atom species with individual spin $f_1=f_2=1$ can be expressed as~\cite{HO1998, OHMI1998}
\begin{equation}
V_{12}(\overrightarrow{r_{1}}-\overrightarrow{r_{2}})=(\alpha+\beta\overrightarrow{f_{1}}\cdot\overrightarrow{f_{2}}+\gamma P_{0})\delta(\overrightarrow{r_{1}}-\overrightarrow{r_{2}}),
\label{eq1}
\end{equation} 
where $\alpha=2 \pi \hbar^2(a_1+a_2)/\mu$, $\beta=2 \pi \hbar^2(a_2-a_1)/\mu$ and $\gamma=2 \pi \hbar^2(2a_0-3a_1+a_2)/\mu$, with $a_F$ the scattering length in total spin channel $F$, $\mu$ the reduced mass, $P_0$ the projection operator to $F = 0$ manifold. $\hbar$ is the reduced Planck constant. Among the three terms, $\beta$ and $\gamma$ are spin-dependent and responsible for the heteronuclear spin dynamics, while the much larger $\alpha$ term is spin-independent and as a result is irrelevant for the current work~\cite{XKLI2015}. 

For the spin-1 mixture, the spin-exchange term $\beta$ is much larger than the singlet pairing term $\gamma$~\cite{ZFXU2009, ZFXU2012}. Thus process~(\ref{eq0}) depends mainly on the $\beta$ term. For the $^{87}$Rb and $^{23}$Na system, we have verified previously that the sign of the $\beta$ term is negative, i.e. ferromagnetic~\cite{XKLI2015}, which tends to align the spins of the two atoms along the same direction.

\begin{figure}
\centering\includegraphics[width=0.9\linewidth]{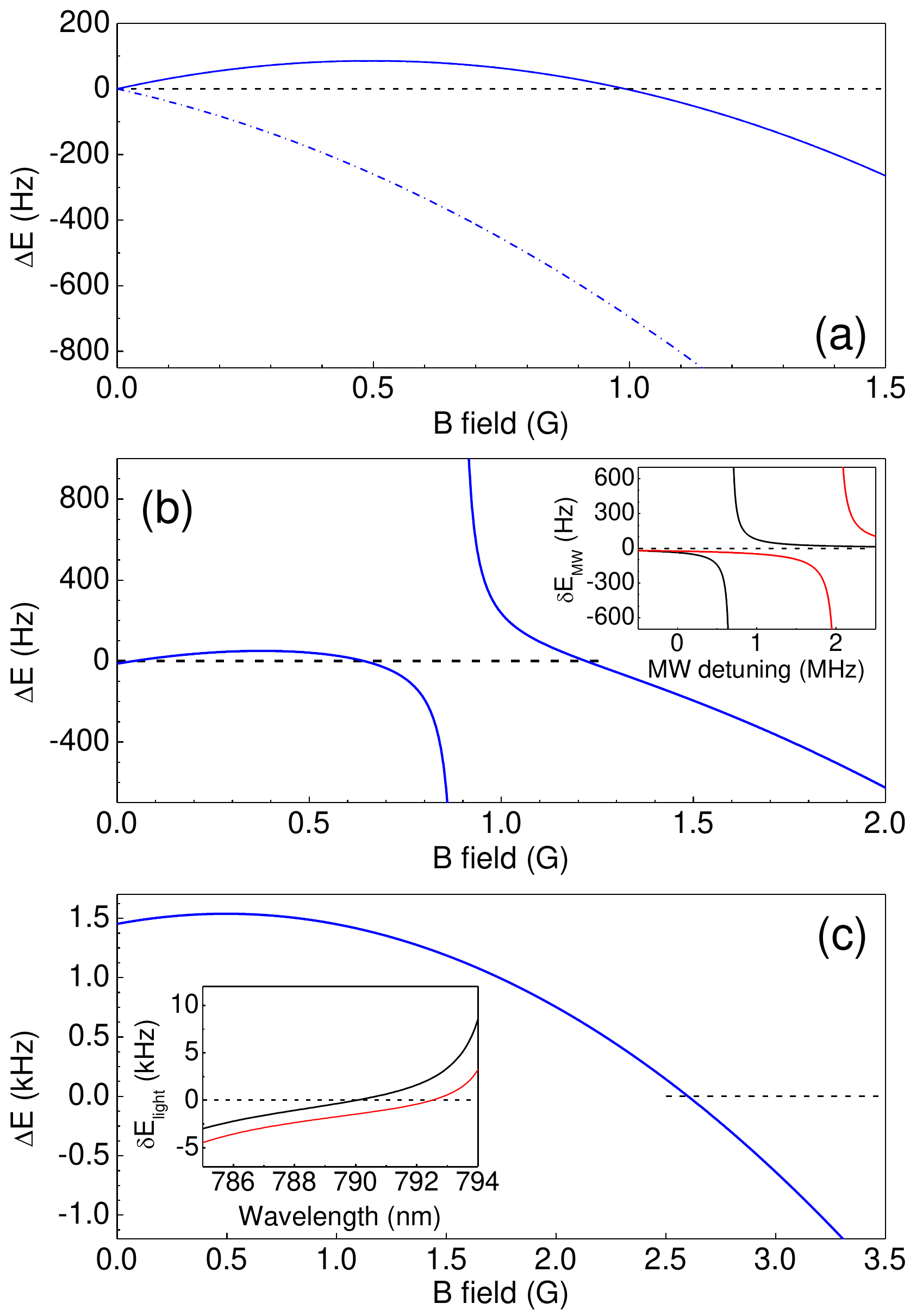}
\caption{$\Delta E$ for the relevant heteronuclear spin oscillation processes in the spin-1 $^{87}$Rb and $^{23}$Na mixture. (a) $\Delta E(B)$ for processes $\ket{0,0} \leftrightarrow \ket{1,-1}$ (solid curve) and $\ket{0,0} \leftrightarrow \ket{-1,1}$ (dash-dot curve). (b) $\Delta E(B,\Delta)$ with the microwave dressing field (Eq.~\ref{eq5}). The microwave is $\sigma^+$ polarized with $\Delta$ = 1.872 MHz. Here, the three non-zero Rabi frequencies are $({\Omega}_{-1,0},{\Omega}_{0,1},{\Omega}_{1,2})=(10 /\sqrt{3},10,10 \sqrt{2})$ kHz. Inset: $\delta E_{\rm MW}(0,\Delta)$ (solid black curve), $\delta E_{MW}(1,\Delta)$ (solid red curve) as a function of $\Delta$ at $B$ = 0.96 G. (c) $\Delta E(B,\Delta_{D1},\Delta_{D2})$ with the vector light shift (Eq.~\ref{eq7}). Here, the intensity of the $\sigma^-$ polarized 790 nm laser is set at 5 W/cm$^2$. Inset: $\delta E_{\rm light}$ for $\ket{0}$ (solid black curve) and  $\ket{1}$ (solid red curve) states of $^{87}$Rb.}
\label{fig1}
\end{figure}

Under the interaction in Eq.~\ref{eq1}, the total magnetization of the system should be conserved. Starting from $\ket{0,0}$ state, both the process (\ref{eq0}) and $\ket{0,0}\leftrightarrow \ket{-1,1}$ can satisfy this requirement. However, these processes are driven by the competition between the $\beta$ term and the total Zeeman energy difference $\Delta E$ between the two sides of the processes. Since the energy scale of the $\beta$ term is only several Hz, the heteronuclear spin oscillation can only occur near the $\Delta E=0$  point. In the $^{87}$Rb and $^{23}$Na system, for process~(\ref{eq0}), 
\begin{equation}
\Delta E(B)=E_{\rm Zeeman}^{\ket{0,0}}-E_{\rm Zeeman}^{\ket{1,-1}}
\label{eq3}
\end{equation}
has a zero crossing at around $B_0 = 0.99$ G as depicted by the solid curve in Fig.~\ref{fig1}(a). Here $E_{\rm Zeeman}^{\ket{m_1,m_2}}$ is the total Zeeman shift of a pair of atoms with $^{87}$Rb in $\ket{m_1}$ and $^{23}$Na in $\ket{m_2}$. Heteronuclear spin exchange following process (1) can thus happen near $B_0$. While for the other process, the magnitude of $\Delta E(B)=E_{\rm Zeeman}^{\ket{0,0}}-E_{\rm Zeeman}^{\ket{-1,1}}$ keeps increasing with $B$ (dash-dot curve in Fig.~\ref{fig1}(a)) and it is thus strongly suppressed near $B_0$.  


We note that homonuclear spinor dynamics for both $^{87}$Rb and $^{23}$Na can also occur accompanying the heteronuclear ones. However, at around 1 G, these dynamics are also largely suppressed by the quadratic Zeeman shifts. Thus working with process~(\ref{eq0}) near $B_0$ will give us both a clean starting point and a very clear signature of heteronuclear dynamics.

\subsection{Tuning $\Delta E$ by microwave}

Microwave dressing is a widely used method for controlling spin oscillation dynamics in homonuclear spinor gases. At a fixed dc magnetic field $B$, the presence of a microwave field off-resonantly connecting the $f=1$ and $f=2$ hyperfine levels can introduce an additional quadratic Zeeman shift. This is due to the differential ac Zeeman shifts $\delta E_{\rm MW}(m,\Delta)$ caused by the microwave because of the differences in detunings and transition strengths for the $m = 0,\pm 1$ Zeeman levels of the $f = 1$ hyperfine state. The sign of the microwave induced quadratic Zeeman shift can also be readily changed by controlling the detuning $\Delta$ for exploring phases not accessible with the dc magnetic field only~\cite{BOOKJANS2011,LCZHAO2014}.

When the Rabi frequency of the microwave transition is much smaller than the detuning $\Delta$, the ac Zeeman shift for each $m$ can be expressed as~\cite{GERBIER2006, LCZHAO2014}  
\begin{equation}
\delta E_{\rm MW}({m},\Delta)=\frac{h}{4}\sum_{m'}{\frac{\Omega^2_{mm'}}{\Delta-(m'g'_f - mg_f)\mu_B/h}}.
\end{equation}   
Here ${\Delta}$ is defined as the detuning with respect to the $\ket{f=1, m=0} \leftrightarrow \ket{f = 2, m'=0}$ transition, ${g}_{f=1} = -0.5018$ and ${g}_{f=2}=0.4998$ are the hyperfine Land\`{e} g-factors, and $\mu_B$ is the Bohr magneton. As shown in the inset of Fig.~\ref{fig1}(b) are $\delta E_{\rm MW}$ for the $m=0$ (solid black curve) and 1 (solid red curve) levels of $f = 1$ $^{87}$Rb atoms at $B = 0.96$ G when $\Delta$ is tuned.

In the heteronuclear spinor system, because of the very different hyperfine splittings, the microwave field can be selectively applied to one species while leave the other species intact. For instance, the hyperfine splitting of $^{87}$Rb is about 6.8 GHz, while that of $^{23}$Na is only around 1.7 GHz. Thus a microwave field near resonance with the $^{87}$Rb hyperfine transition will not affect the energy levels of $^{23}$Na because of the large detuning. This can still be used to control the spin dynamics in process~(\ref{eq0}) since $\delta E_{\rm MW}$ is different for $\ket{f=1, m=0}$ and $\ket{f=1, m=1}$ levels of $^{87}$Rb.

Taking the microwave induced shifts on the $^{87}$Rb energy levels into account, the total internal energy difference between the two sides of process~(\ref{eq0}) is now    
\begin{equation}
\begin{split}
\Delta E(B,\Delta)=(E_{\rm Zeeman}^{\ket{0,0}} -E_{\rm Zeeman}^{\ket{1,-1}}) \\
 + (\delta E_{\rm MW}({0},\Delta) -\delta E_{\rm MW}({1},\Delta)).
\end{split}
\label{eq5}
\end{equation}
At a fixed $B$, by changing the detuning $\Delta$ of the microwave field, $\Delta E$ and thus the spin dynamics can be tuned. 

As shown in Fig.~\ref{fig1}(b), applying a $\sigma^+$ polarized microwave field with fixed frequency and power while varying $B$, the zero crossing point of $\Delta E$ will be shifted to different $B$ field values. For this calculation, the microwave frequency is 6836.5545 MHz while $B$ is tuned from 0 to 2 G. Due to selection rules, there are only three non-zero microwave Rabi frequencies (${\Omega}_{-1,0},{\Omega}_{0,1},{\Omega}_{1,2}$) with ratios determined by the relative transition strengths. We note that under this configuration, there are actually several zero crossings in Fig.~\ref{fig1}(b) due to the Zeeman levels of the $f=2$ hyperfine state. In principle, all of them can be used for spin dynamics control.




\subsection{Tuning $\Delta E$ by vector light shift}

In ref.~\cite{XKLI2015}, we already demonstrated controlling the resonance position of the heteronuclear spin oscillations with species and spin-dependent vector light shift. In that work, the vector light shift was induced by adding various amounts of circular polarization components to the optical trap laser with a quarter waveplate. Here we introduce an additional single frequency laser to induce the vector light shift more flexibly.      

For alkali atoms, when the laser detuning is much larger than the excited state hyperfine splittings, the light shift can be generally expressed as~\cite{GRIMM2000}
\begin{equation}
\begin{split}
\delta E_{\rm light}(m,\Delta_{D1},\Delta_{D2}) = \frac{\pi c^2 \Gamma}{2 \omega^3_0}\left[\left(\frac{2}{\Delta_{D2}} + \frac{1}{\Delta_{D1}}\right)\right. \\
+\left.\wp\left(\frac{ g_f m}{\Delta_{D2}} - \frac{ g_f m}{\Delta_{D1}}\right)\right] I(\vec{r}).
\end{split}
\label{eq6}
\end{equation}
Here $\Gamma$ is the linewidth of the $D-$lines, $\omega_0$ is the transition frequency, $\Delta_{D1}$ ($\Delta_{D2}$) is the detuning of the laser with respect to the $D_1$ ($D_2$) line, and $\wp = 0$ and $\pm1$ for linear and circular $\sigma^{\pm}$ polarized light. In the above equation, the first term comes from the spin-independent scalar ac polarizibility. If the laser frequency is tuned to in between the excited-state fine structures, this part of the light shift could become zero as the signs of $\Delta_{D1}$ and $\Delta_{D1}$ are opposite. For $^{87}$Rb, the corresponding wavelength for zero scalar light shift is 790.0 nm. The second, spin-dependent term, which is only non-zero when circular polarized light is used ($\wp\neq 0$), is from the vector polarizibility. The inset of Fig.~\ref{fig1}(c) shows $\delta E_{\rm light}$ vs. the dressing laser wavelength for $\ket{0}$ (solid black curve) and $\ket{1}$ (solid red curve) Zeeman levels of $^{87}$Rb under low magnetic field. The light polarization is $\sigma^-$ ($\wp = -1$).      

In the heteronuclear $^{87}$Rb and $^{23}$Na spinor system, since the $D$-line transition frequencies are very different for the two species and $\delta E_{\rm light}$ is inversely proportional to the detuning, the light shift can also be made essentially species-selective. In the experiment, we use a laser operating at around 790 nm which affects mainly the energy levels of $^{87}$Rb with negligible effect on $^{23}$Na. 

Taking $\delta E_{\rm light}$ on $^{87}$Rb into account, the total internal energy difference between the two sides of process~(\ref{eq0}) can be expressed as    
\begin{equation}
\begin{split}
\Delta E(B,\Delta_{D1},\Delta_{D2})=(E_{\rm Zeeman}^{\ket{0,0}} -E_{\rm Zeeman}^{\ket{1,-1}}) \\
 + [\delta E_{\rm light}({0},\Delta_{D1},\Delta_{D2}) -\delta E_{\rm light}({1},\Delta_{D1},\Delta_{D2})].
\end{split}
\label{eq7}
\end{equation}
It can be seen that only the spin-dependent vector light shift has influence on the spin dynamics. We note that the use of near 790 nm light also minimizes possible perturbation to the optical trap potential which otherwise will modify the sample density distribution. As shown in Fig.~\ref{fig1}(c), fixing the laser wavelength and intensity while changing the $B$ fields, the zero crossing point of $\Delta E$ can be tuned to far from that without the $\sigma^-$ polarized laser. 


In the homonuclear case, since the vector light shifts of the $\ket{+1}$ and $\ket{-1}$ spin states have the same magnitude but the opposite sign, the process $2\ket{0} \leftrightarrow \ket{+1} + \ket{-1}$ is not sensitive to uniform light field illumination.


\section{Experiment and results}
\label{experiment}

\subsection{Spinor mixture preparation and spin dynamics detection} 
\label{setup}

We produce the ultracold $^{87}$Rb and $^{23}$Na mixture in a crossed optical trap formed by two linearly polarized 1070 nm laser beams with both atoms in their $\ket{-1}$ spin state. By adjusting the final evaporation in the optical trap, either a mixture of essentially pure Bose-Einstein condensate (BEC) of both species or a pure $^{23}$Na BEC plus a $^{87}$Rb thermal cloud can be obtained. The magnetic field is then ramped up to 60 G and subsequently a single rapid adiabatic passage is applied to transfer simultaneously both atoms to their $\ket{0}$ state with near 100\% efficiency. Such a high $B$ filed is necessary to generate enough frequency differences between the $\ket{-1}\rightarrow\ket{0}$ and $\ket{0}\rightarrow\ket{1}$ transitions in order to avoid populating the $\ket{1}$ states via the cascade transition $\ket{-1}\rightarrow\ket{0}\rightarrow\ket{1}$. 

The $\ket{0,0}$ mixture is then hold at 4 G for one second to make sure full equilibrium is reached. At this stage, no spin dynamics is detected since $\Delta E$ is high. Finally, the magnetic field is ramped to a range of lower values to observe the spin population oscillations at different holding time $t$. For detection, we switch off the optical trap and apply a magnetic field gradient to separate the different spin states during the time of flight expansion. Absorption image is then used to record the number of atoms $N_m^i$ in each spin states $\ket{m}$ of species $i$, with $i$ = Na or Rb. The fractional spin population $\rho_m^i = N_m^i/N^i$ can then be obtained from the total number of each species $N^i = N^i_{-1} + N^i_{0} + N^i_{+1}$.


\subsection{Coherent heteronuclear spin oscillations in the double BEC mixture}
\label{result1}

\begin{figure}[t]
\centering\includegraphics[width=0.9\linewidth]{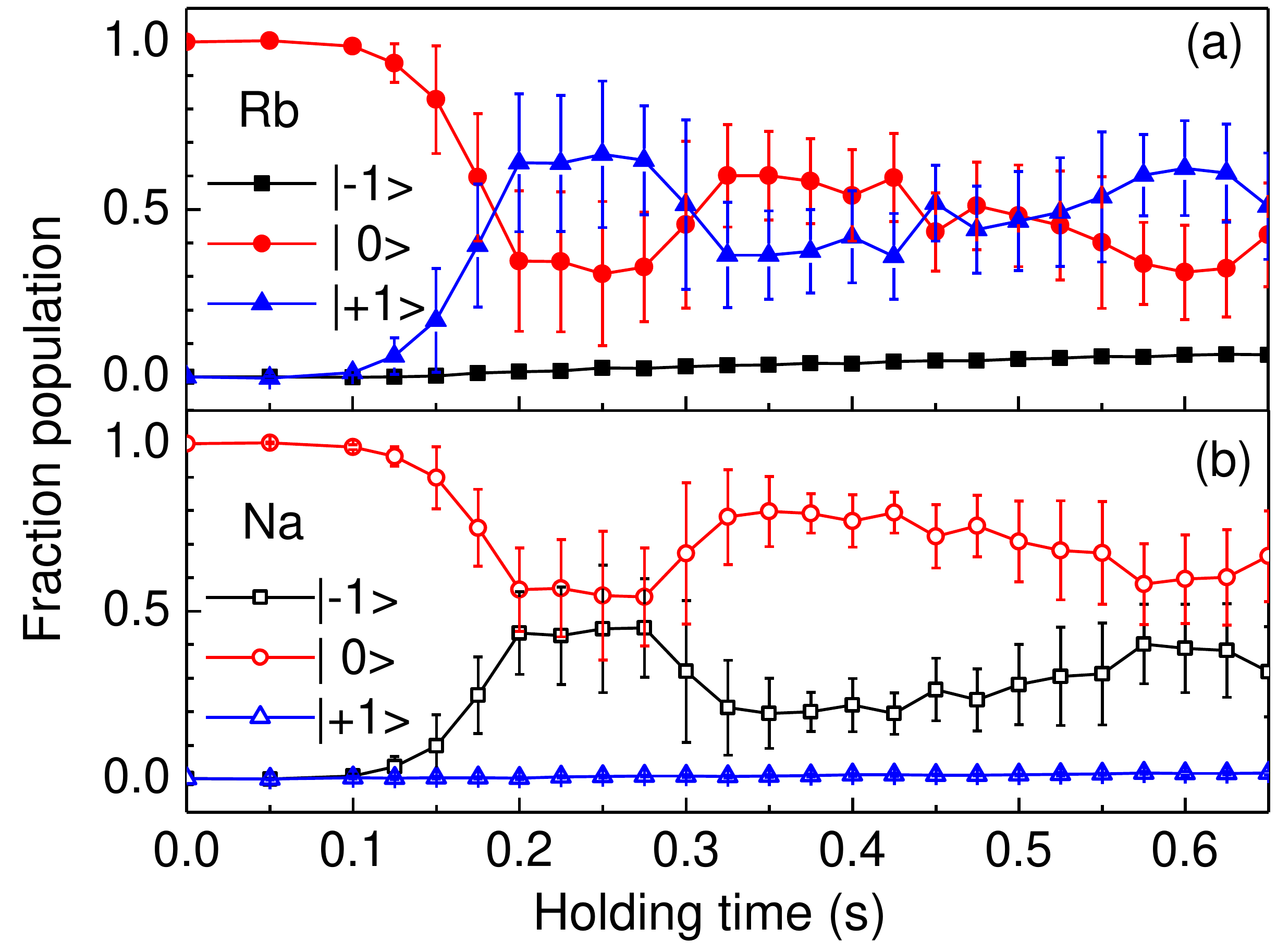}
\caption{Coherent heteronuclear spin oscillations in the spin-1 $^{87}$Rb and $^{23}$Na dual BEC mixture following the $\ket{0,0} \leftrightarrow \ket{1,-1}$ process. (a) and (b) show the time evolution of the spin population for $^{87}$Rb and $^{23}$Na, respectively. The magnetic field is held at $B$ = 0.922 G during the whole process. The measured trap frequencies are $(\omega_x,\omega_y,\omega_z) = 2\pi\times(240,240,120)$ Hz for Rb and $2\pi\times(280,280,140)$ Hz for Na. The number of atoms in the Rb (Na) condensate is $3.0(2)\times 10^4$ ($5.8(4)\times10^4$). The calculated peak densities are $3.0\times 10^{14}$ cm$^{-3}$ and $1.4\times 10^{14}$ cm$^{-3}$ for Rb and Na, respectively. Error bars represent 1 standard deviation of typically six repetitions.} 
\label{fig2}
\end{figure}

We have investigated the heteronuclear spin dynamics in the double BEC mixture. Fig.~\ref{fig2} shows the coherent spin population following process~(\ref{eq0}) measured with the double BEC at $B = 0.922$ G. Similar to the previously investigated $\ket{0, -1} \leftrightarrow \ket{-1, 0}$ case~\citep{XKLI2015}, the synchronized oscillations between the two species as well as between different components of the same species are obvious signatures of the coherent heteronuclear spin dynamics.

With both the double BEC mixture and the BEC + thermal mixture, we have observed the appearance of the third spin state for both $^{87}$Rb and $^{23}$Na (Fig.2). We have verified experimentally that, with either $^{87}$Rb or $^{23}$Na atoms alone in spin state $\ket{0}$, no homonuclear spin dynamics can happen at the range of magnetic fields used in this investigation. We thus believe these small amount of population is a result of other heteronuclear spin processes. For instance, the $\ket{-1}$ $^{23}$Na atoms generated by process~(\ref{eq0}) may initiate the $\ket{0, -1} \leftrightarrow \ket{-1, 0}$ process to produce $^{87}$Rb atom in $\ket{-1}$ spin state.

In the previous work, the starting point is a non-equilibrium spin configuration, thus the $\ket{0, -1} \leftrightarrow \ket{-1, 0}$ spin population oscillation always happens immediately~\cite{XKLI2015}. Here, starting from the zero magnetization $\ket{0,0}$ state, a delay can be observed before the spin dynamics starts. In Fig.~\ref{fig2}, this delay is on the order of 150 ms. This is very similar to the homonuclear spin-1 case, i.e., starting from spin state $\ket{0}$ of $^{87}$Rb or $^{23}$Na, the homonuclear spin population oscillation $2\ket{0}\leftrightarrow \ket{-1} + \ket{1}$ always starts after a delay~\cite{CHANG2004,BOOKJANS2011}. Such a behavior can be explained by the fact that the $\ket{0}$ state of the homonuclear spin-1 system is metastable and spin dynamics can only be initiated by quantum fluctuation which needs some time to build up. Similar physics may be also dictating the heteronuclear dynamics here, but this has not been investigated thoroughly.   


\subsection{Spin dynamics in the mixture of a $^{87}$Rb thermal cloud and a $^{23}$Na BEC}
\label{result2}

\begin{figure}
\centering\includegraphics[width=0.9\linewidth]{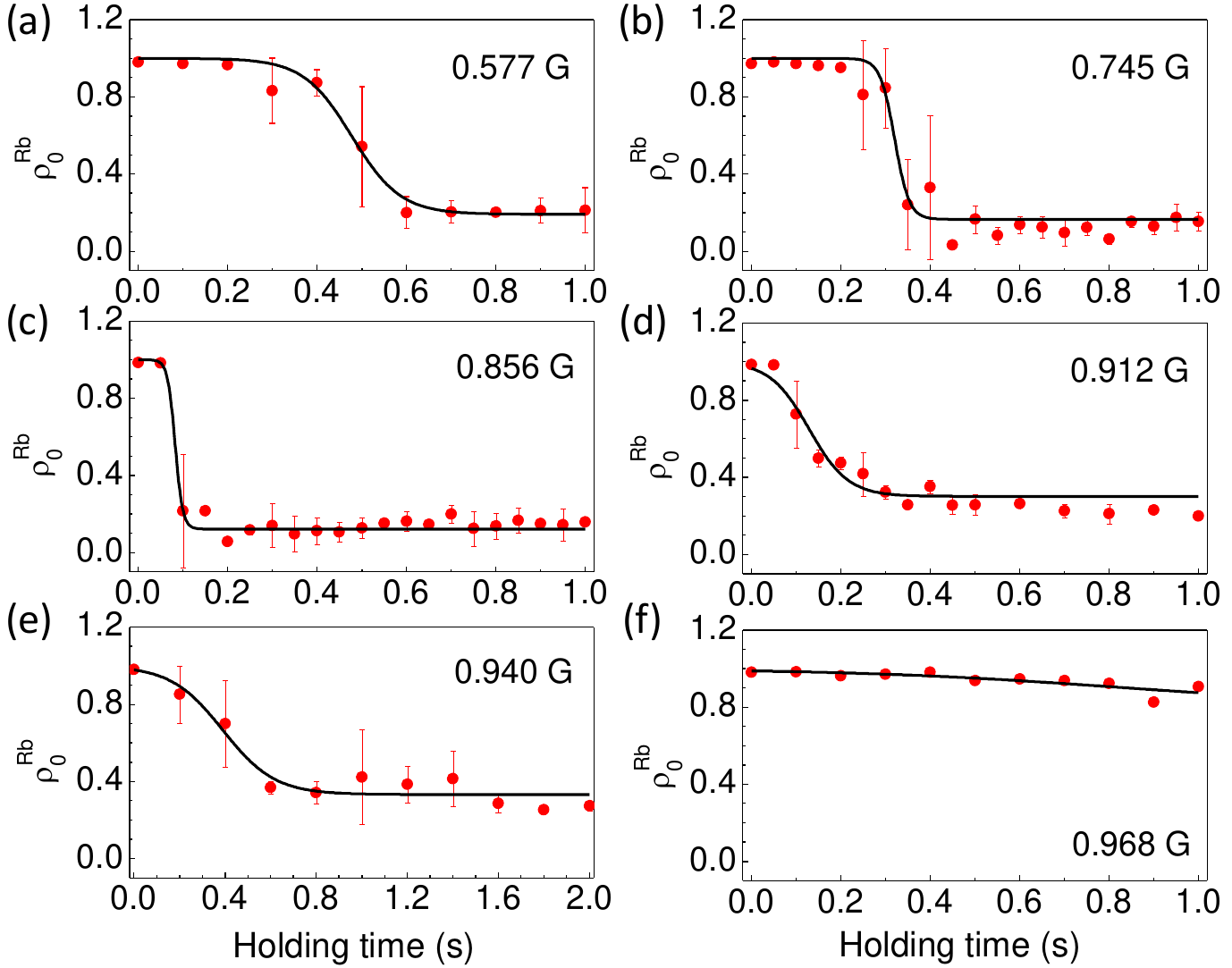}
\caption{Heteronuclear spin dynamics in the thermal $^{87}$Rb and $^{23}$Na BEC spin-1 mixture following process~(\ref{eq0}). Time evolution of $\rho_0^{\rm Rb}$ for several magnetic fields. The solid curves are from fit to the Sigmoid function. The error bars represent 1 standard deviation of typically 3 to 5 repetitions. For these measurements, a nearly spherical trap with measured trap frequencies of $2\pi\times64$ Hz for Rb and $2\pi\times72$ Hz for Na is used. The calculated peak densities are $4.6\times 10^{12}$ cm$^{-3}$ for the $^{87}$Rb thermal gas and $5.6\times 10^{13}$ cm$^{-3}$ for the $^{23}$Na BEC. }       
\label{fig3}
\end{figure}

A complication in using the double $^{87}$Rb and $^{23}$Na BEC mixture is the phase separation which leads to poor density overlap. This is due to the relatively large interspecies repulsive interaction~\cite{FDWANG2015}. In addition, the difference in the trap frequencies for the two species results in a differential gravitational sag which displaces the centers of mass of the two clouds in the vertical direction. The rather high trap frequencies used in Fig.~\ref{fig2} are chosen to compensate these effects and increase the density overlap. Nevertheless, we have found experimentally that the spin dynamics in the BEC mixture depends very sensitively on the optical dipole trap. For instance, when a very weak optical trap is used, the spin oscillation becomes totally non-repeatable. We believe this is because of the modulation of the double BEC overlap and thus the spin exchange energy due to the aforementioned reasons. Because of this, for the purpose of investigating control of the spin dynamics, we choose to use a mixture of a thermal cloud of $^{87}$Rb atoms and a BEC of $^{23}$Na for the rest of this work. In this configuration, the overlap between the two species is always good, and the signal is more repeatable. 

\begin{figure}
\centering\includegraphics[width=0.9\linewidth]{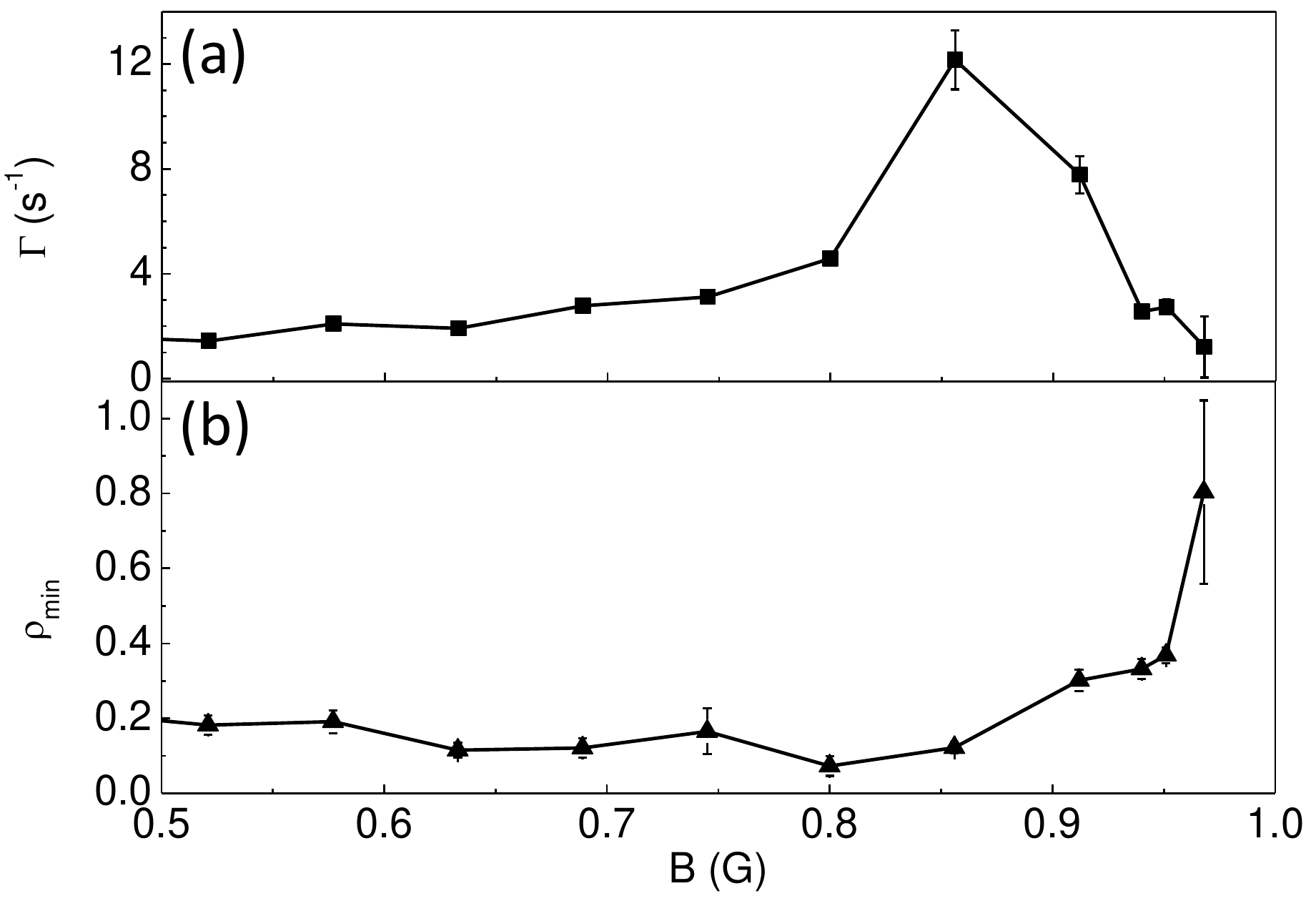}
\caption{Spin exchange rate $\Gamma$ and final population $\rho_{\rm min}$ extracted from Fig.~\ref{fig3}. (a). $\Gamma$ versus magnetic field extracted from the Sigmoid fitting. (b). $\rho_{\rm min}$ versus magnetic field from the Sigmoid fitting. The curves are for eye guiding.}       
\label{fig4}
\end{figure}

Similar to the double BEC case, correlated heteronuclear spin population changes following process~(\ref{eq0}) can also be observed. However, as shown by the time evolution of $\rho_0^{\rm Rb}$ in Fig.~\ref{fig3}, the oscillations are strongly over-damped with no periodical features. From Fig.~\ref{fig3}(a)--(f), the $B$ field dependence can be clearly observed from changes of the delay time before the dynamics and the equilibrium fractional population. To quantify this dependence, we fit the data with a Sigmoid function to extract both the cross-over time $T_{1/2}$ and the final saturated fractional population $\rho_{\rm min}$ of $^{87}$Rb, with the relation between the two parameters defined by $\rho_0^{\rm Rb}(T_{1/2})=(1 +\rho_{\rm min})/2$~\cite{BOOKJANS2011}. Fig.~\ref{fig4}(a) and (b) show the rate of the spin exchange $\Gamma= 1/T_{1/2}$ and $\rho_{\rm min}$ versus $B$. A maximum of $\Gamma$ can be observed at around 0.85 G, beyond which the spin dynamics slows down and $\rho_{\rm min}$ keeps increasing. Near the zero crossing point of $\Delta E$ at 0.99 G, $\rho_{\rm min}$ is about 0.8 after one second which indicates a very slow spin dynamics. 

Currently, we lack a quantitative understanding of the mismatch between the maximum of $\Gamma$ and the zero crossing of $\Delta E$. The fact that we are using a mixture of BEC and thermal atoms makes it hard to treat the problem theoretically, especially with the possible thermal and quantum fluctuations involved. Nevertheless, as has been pointed out in~\cite{XKLI2015}, the spinor mixture is a many-body system, while the intuitive understanding based on the argument of $\Delta E$ is only true for two particles. The exact peak position of $\Gamma$ also depends on the homonuclear spin-dependent interactions and possibly the number ratios.

\subsection{Tuning spin dynamics with microwave}
\label{result3}

\begin{figure}
\centering\includegraphics[width=0.9\linewidth]{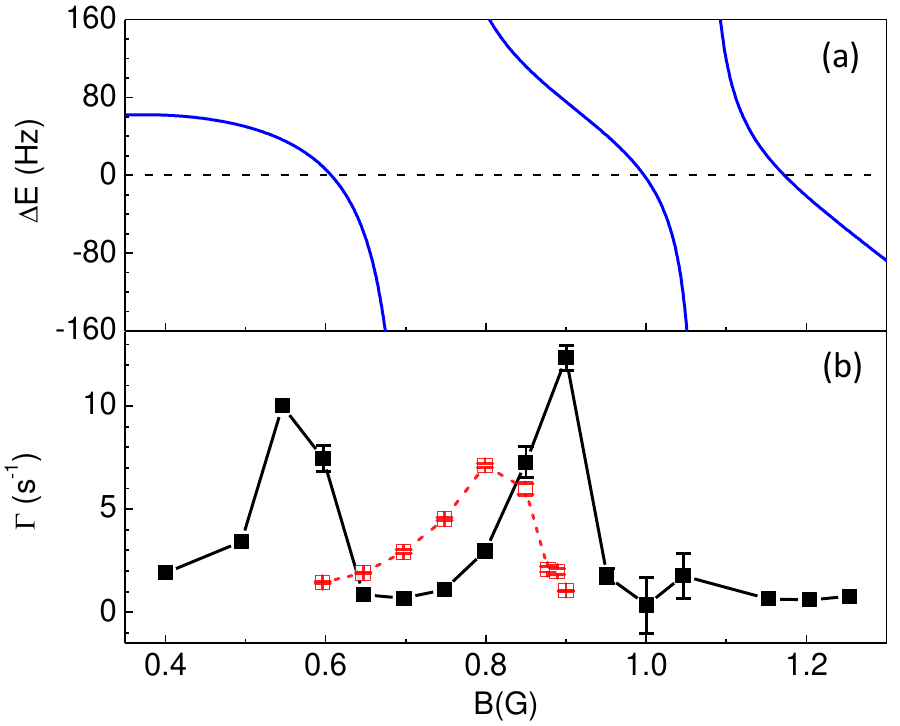}
\caption{Control of heteronuclear spin dynamics with microwave. (a). The blue curve shows $\Delta E(B,\Delta)$ with the microwave dressing field, which is calculated with Eq.~\ref{eq5} using experimentally measured parameters. (b). The black solid squares are the measured $\Gamma$ of the spin dynamics in presence of the microwave; the red open squares are the case without the microwave. All measurements are performed at the same atomic conditions with peak densities of $2.2\times 10^{12}$~cm$^{-3}$ for the $^{87}$Rb thermal cloud and $5.0\times 10^{13}$~cm$^{-3}$ for the $^{23}$Na BEC. The curves are for eye guiding.}      
\label{fig5}
\end{figure}

To demonstrate microwave control of the heteronuclear spin dynamics, we broadcast a microwave signal to the atoms after the $B$ field is ramped to the final value. In our experiment, the polarization of the microwave field is not well controlled and the exact amount of power radiated on the atoms is hard to measure directly. To calibrate the microwave field, we drive Rabi oscillations of several $\pi$ and $\sigma^{\pm}$ transitions between the $f = 1$ and $f = 2$ hyperfine Zeeman levels of $^{87}$Rb. From the measured Rabi frequencies of these transitions, we deduce the polarization distribution and intensity of the microwave field using the well-known relative transition strengths. Rabi frequencies of all other relevant transitions can then be obtained accurately. 

In the experiment, the microwave frequency is fixed at 6836.18 MHz. For the microwave power used in this work, the several measured on resonance Rabi frequencies are ${\Omega}_{-1,-2}=8.1$ kHz, ${\Omega}_{-1,-1}=4.5$ kHz and ${\Omega}_{-1,0}=3.7$ kHz, i.e., the microwave field has a mixed polarization which can drive all possible transitions. We then calculate the other several Rabi frequencies as ${\Omega}_{0,-1}=5.7$ kHz, ${\Omega}_{0,0}=5.1$ kHz, ${\Omega}_{0,1}=6.3$ kHz ${\Omega}_{1,0}=3.3$ kHz, ${\Omega}_{1,1}=4.5$ kHz and ${\Omega}_{1,2}=8.9$ kHz. Fig.~\ref{fig5}(a) shows $\Delta E$ versus magnetic field calculated from Eq.~\ref{eq5} with the calibrated on resonant Rabi frequencies. The three zero crossings are all within the $B$ field range for observing the heteronuclear spin dynamics.

We map out the $B$ field dependence of the spin dynamics in presence of the microwave dressing by measuring the rate of the spin exchange $\Gamma$ following the same procedure as in section~\ref{experiment}.C. We note that the peak position of $\Gamma$ depends on the atomic conditions due to the resulting change of the spin dependent interaction. To see a clear signature of the microwave dressing effect, it is thus important to perform all the measurements with the same atomic conditions. As shown in Fig.~\ref{fig5}(b), without the microwave dressing, the peak of $\Gamma$ appears at about 0.80 G. At the same atomic conditions but with the microwave dressing, two peaks of $\Gamma$ can be observed at around 0.55 G and 0.88 G, which are close to the two zero crossings of $\Delta E$ at 0.6 G and 1 G, respectively. Same as the case without the additional dressing fields, the maximums of $\Gamma$ also occur at $B$ fields lower than the $\Delta E=0$ points. 

The apparent shifts of the peak positions of $\Gamma$ and the appearance of the additional peak are both clear manifestation of manipulations of the heteronuclear spin dynamics with microwave. However, no peak of $\Gamma$ is observed for the zero crossing at around 1.2 G. The reason behind this is not fully understood.

\subsection{Tuning spin dynamics with vector light shift}
\label{result4}

\begin{figure}
\centering\includegraphics[width=0.8\linewidth]{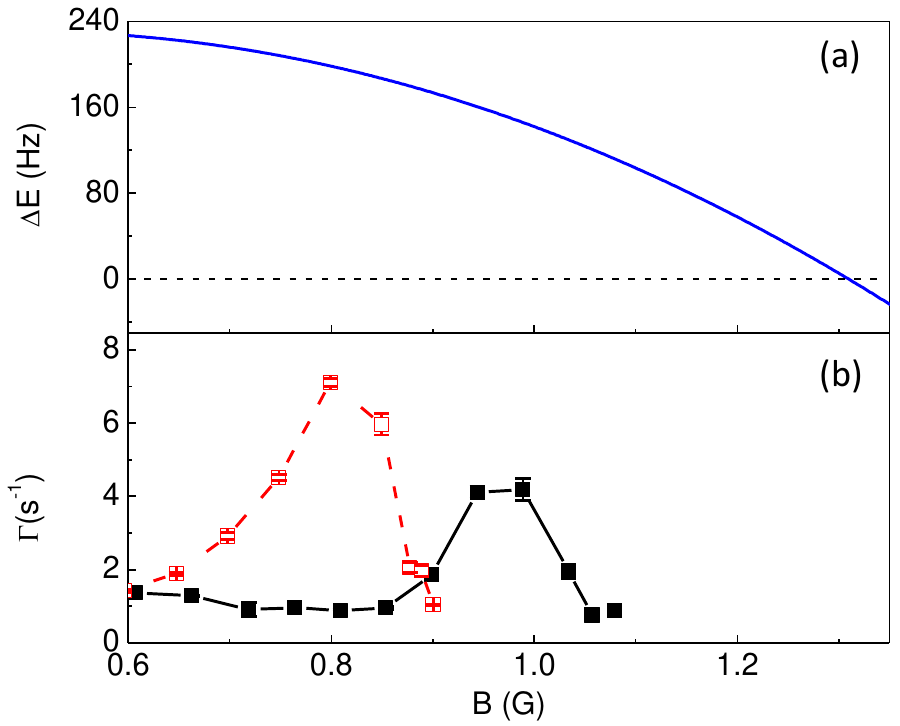}
\caption{Control of heteronuclear spin dynamics with vector light shift. (a) The blue curve shows the calculated $\Delta E(B,\Delta_{D1},\Delta_{D2})$ from Eq.~\ref{eq7} with the $\sigma^-$ polarized 789.817 nm laser beam. The calculation is based on the peak intensity of the laser beam. (b) The black solid squares are the measured $\Gamma$ with the vector light shift. The atomic conditions are the same as those used in Fig.~\ref{fig5}. For comparison, the red open squares show the case without the laser beam. The curves are for eye guiding. }      
\label{fig6}
\end{figure}

As discussed in sec.~\ref{theory}(c), using light field to tune spin dynamics is a unique capability in the heteronuclear spinor system. Since process~(\ref{eq0}) depends directly on the linear Zeeman shift, the effective magnetic field generated by the vector light shift can induce a large change to the spin oscillation resonance. Experimentally, we verify this with the help of a free running external cavity diode laser tuned to 789.817 nm. It is introduced to the atoms along the quantization axis defined by the magnetic field after passing through a $\lambda/4$ waveplate for obtaining $\sigma^-$ polarization. The 1.15 mm beam diameter is much larger than the sizes of the atomic clouds to ensure a uniform illumination. From the measured laser power, the peak intensity is calculated to be 0.5 $\rm W/cm^2$ which is enough to induce a differential vector light shift of $2\pi \times 140$~Hz between the $\ket{0}$ and $\ket{1}$ states of $^{87}$Rb. At this intensity, no significant shortening of the trapping lifetime is observed. As depicted in Fig.~\ref{fig6}(a), in this configuration the calculated zero crossing point of $\Delta E$ with Eq.~\ref{eq7} is shifted to 1.31 G.

Figure~\ref{fig6}(b) shows $\Gamma$ extracted from measuring the spin dynamics at different $B$ field with the light beam. The measurement was performed with essentially the same atomic condition as Fig.~\ref{fig5}. Comparing with the case without the light field (red open squares), the measured shift of the resonance is 0.17 G. This is about 50\% smaller than the shift of the zero-crossing points in $\Delta E$. This disagreement could be due to imperfections in the laser beam polarization and/or alignment which result in a smaller shift than the calculation. 

It is also notice that the peak of $\Gamma$ is significantly smaller than the B field only case. One possible reason for this is the laser power fluctuation. Since $\Delta E$ depends on the laser power, such fluctuation may diminish the spin population oscillation. In future experiments, the laser power should be carefully stabilized with a feedback control system.



\section{Conclusion}

In summary, we have observed and developed controlling methods for heteronuclear coherent spin dynamics. With the process $\ket{0,0} \leftrightarrow \ket{1,-1}$ in the spin-1 $^{87}$Rb and $^{23}$Na mixture, we showed that both the detuned microwave field and circular polarized light field can tune spin dynamics to occur in regions not accessible with magnetic field only. Since both microwave and light fields can be controlled in fast time scale, the methods studied here should be useful for tuning the spin dynamics time-dependently, such as quenching. In addition, we find the versatile spatial control of light could be a valuable capability. For instance, by shrinking the laser beam size to smaller than that of the  atomic sample, it is possible to introduce local spin dynamics manipulation~\cite{LESLIE200902, CHOI2012}. It is also possible to form a standing wave to induce a periodical modulation to the spin dynamics.

\begin{acknowledgments}
We thank Qi Zhou and Peng Zhang for valuable discussions. This work was supported by the Hong Kong Research Grants Council GRF CUHK 14305214 and 14304015, HKU 17303215 and CRF C6026-16W.
\end{acknowledgments}

\end{document}